\documentclass[12pt,preprint]{aastex}

\shorttitle{A Stellar Stream Behind NGC 188}

\begin{document}

\renewcommand{\thefootnote}{\fnsymbol{footnote}}
\setcounter{footnote}{1}

\title{Detection of a Stellar Stream Behind Open Cluster NGC 188: Another Part of the Monoceros 
Stream\footnote{Paper 42 of the WIYN Open Cluster Study (WOCS).}}

\author{Dana I. Casetti-Dinescu\altaffilmark{1,2}, Terrence M. Girard\altaffilmark{1}, 
Imants Platais\altaffilmark{3} and William F. van Altena\altaffilmark{1}}

\altaffiltext{1}{Astronomy Department, Yale University, P.O. Box 208101,
New Haven, CT 06520-8101, USA}
\altaffiltext{2}{Astronomical Institute of the Romanian Academy, Str.
Cutitul de Argint 5, RO-75212, Bucharest 28, Romania}
\altaffiltext{3}{Department of Physics and Astronomy, Johns Hopkins University, 3400 North Charles Street,
Baltimore, MD 21218}

\email{dana.casetti@yale.edu}

\begin{abstract}

We present results from a 
WIYN/OPTIC photometric and astrometric survey of the field of the open cluster 
NGC 188 ((l,b) = (122.8\arcdeg, 22.5\arcdeg)). 
We combine these results with the 
proper-motion and photometry catalog  of Platais et al.  and
demonstrate the existence of a stellar overdensity in the background of NGC 188.
The theoretical isochrone fits to the color-magnitude diagram of the overdensity are
consistent with an age between 6 and 10 Gyr and an intermediately metal poor population ([Fe/H] = -0.5 to -1.0). 
The distance to the overdensity is estimated to be between 10.0 and 12.6 kpc. 
The proper-motions indicate that the stellar population of the overdensity is kinematically cold.

The distance estimate and the absolute proper motion of the overdensity 
agree reasonably well with the predictions of the Pe\~{n}arrubia et al.  model of the 
formation of the Monoceros stream. Orbits for this material constructed with plausible
radial-velocity values, indicate that dynamically, this material is unlikely to 
belong to the thick disk.
Taken together, this evidence suggests that the
newly-found overdensity is part of the Monoceros stream.

\end{abstract}


\section{Introduction}

In the last decade, large-scale surveys have demonstrated the existence of 
extended tidal streams and overdensities that are believed to be 
generated by the disruption of massive systems such as dwarf galaxies as they merge with our Galaxy.
To date, there are three reasonably well-documented large structures known to reside in the halo and outer disk of the Galaxy:
the Sagittarius stream  with its known progenitor the Sagittarius dwarf galaxy
(Ibata et al. 1994, Ibata et al. 2001,  Newberg et al. 2002, Majewski et al. 2003), the Monoceros stream
(Newberg et al. 2002, Yanny et al. 2003, Ibata et al. 2003, Rocha-Pinto et al. 2003, Conn et al. 2005, 2007, 2008)
with the much-disputed possible progenitor the overdensity in Canis Major, and finally the 
Virgo Overdensity (Newberg et al. 2002, Vivas \& Zinn 2003, Juri\'{c} et al. 2008) with no known 
progenitor (see however Casetti-Dinescu et al. 2009 for a tentative suggestion). These 
structures' formation has direct implications on the history of the assembling of the Milky Way and 
thus on the merger history in a $\Lambda$CDM cosmology.

\renewcommand{\thefootnote}{\arabic{footnote}}
\setcounter{footnote}{0}

Here we investigate a stellar overdensity found in the background of the open cluster NGC 188, and
provide evidence that is is part of the Monoceros stream. Our data may help better constrain models
of the formation and origin of this immense ring-like feature mapped from l = $60\arcdeg$ to $300\arcdeg$, above and
below the Galactic plane (Conn et al. 2008 and references therein). In this investigation, 
we make use of the excellent absolute proper-motions and photometry catalog provided by Platais et al. (2003, hereafter P03)
over an area of 0.75 square degrees down to $V = 21$. We also add new, deep $VI$ photometry for an area half that of P03
obtained with WIYN 3.5m\footnote{The WIYN Observatory is a joint facility of the University of Wisconsin-Madison, Indiana University,
Yale University and the National Optical Astronomy Observatory.}, that reaches $V \sim 23.5$.

We present our data in Section 2, the color-magnitude diagram (CMD) interpretation in Section 3, the proper-motion 
analysis in Section 4, and discuss the origin of this newly-found overdensity in Section 5. In Section 6
we summarize our results.

\section{The Data}

In this work we use the $BV$ photometry and proper-motion data from P03, and the
$VI$ photometry obtained during a 3-night run in October 2007 at WIYN with the 
Orthogonal Parallel Transfer Imaging Camera (OPTIC). The WIYN data were designed to
cover a large area in the cluster region with the intent to: 1) better characterize the 
stellar overdensity found in the P03 data, and 2) combine with a previous WIYN/OPTIC dataset
taken in 2003 to characterize the astrometric properties and the proper-motion precision
that can be obtained with this set-up. The astrometric aspect of the WIYN/OPTIC data set
will be the subject of another paper, as part of a study that aims to characterize 
a similar type of detector with an area coverage of one square degree to be installed 
on WIYN early in 2010 and known as the One Degree Imager (ODI, Jacoby et al. 2002).

In Figure 1 we show the area coverage of the P03 catalog (the round area), 
our WIYN/OPTIC dataset (in gray), and the $VI$ calibration field which is the data from 
von Hippel \& Sarajedini (1998) (in black). Our data set covers $53\%$ of the P03 area.

\subsection{The Platais et al. Catalog}

The P03 data include calibrated CCD $BV$ photometry and proper-motions down to a limiting 
magnitude of $V \sim 21$ for an area of 0.75 square degrees.
A star/galaxy classification is also provided which lists 100 galaxies.
The proper motions were derived from a combination of photographic plates and
 KPNO 4m Mosaic Imager data spanning $~40$ years.
Proper-motion uncertainties are 0.15 mas~yr$^{-1}$ for well-measured objects, i.e., 
to $V\sim 15$, and below 1.0 mas~yr$^{-1}$ down to $V \sim 20$. 
The CMD from the P03 photometry clearly shows a main sequence-like
feature (see Fig 8 in P03) between $V = 19$ and 21, and $B-V = 0.4$ to 0.8. This feature
stimulated us to obtain a deeper, large-area data set.  Other deep photometric studies 
of NGC~188 have not captured this feature, as they were concentrated on 
small areas on the cluster. Studies with somewhat larger areas are too shallow to sample 
the feature.

\subsection{The WIYN/OPTIC Data}

We have imaged an area of $\sim 40\arcmin\times 40\arcmin$ around NGC~188 using 
overlapping OPTIC frames.
This camera is able to compensate for real-time image motion providing
tip/tilt corrections without additional optics or moving parts 
via orthogonal transfer shifting of the charge (Tonry et al. 1997) thus
effectively improving the seeing. It consists of two 2K by 4K
orthogonal transfer CCDs  mounted adjacent to each other in a single dewar. Each 
CCD is electronically divided into two regions, and each of these regions has two parts: 
a guide region and a science region. The guide regions are at the top and 
bottom of each CCD and are $2048 \times 516$ pixels. When the orthogonal transfer shift 
is not engaged, OPTIC functions as a conventional imager with the guide regions
used as science regions. Amplifiers are placed at the 
end of each chip, therefore the detector conceptually is a square array of four 2K by 2K CCDs.
At WIYN, the pixel scale is $0.14\arcsec$, thus covering an area of 
$9.6\arcmin\times9.6\arcmin$. It has a read noise of 4 electrons, a nominal 
gain of 1.4 e/ADU, and a readout time of 25 sec. 

The OPTIC frames were designed to have a center-to-corner overlap, centered on 
the cluster, and covering as much of the P03 data set as possible. For each frame
we have typically taken a 600-sec exposure in Harris $V$ and a 300-sec exposure in 
Harris $I$.
One frame at the center of our area had three repeated exposures in each filter,
each of the same nominal duration as for the typical observations. Hereafter we will call this 
frame, the ``central-deep field''. The seeing varied
between 0.8 to 2 arcsec (in the $I$ band) during the entire run, 
therefore our data varies in quality and depth. The central-deep field was observed in
1.0\arcsec seeing. Since the purpose of this study was to cover as large an area as
possible to match the P03 set, and given that the seeing was rather poor for most of the
data, we have used OPTIC without the OT engaged for the entire run.

Each OPTIC frame was divided into its four amplifier subunits, and each subunit was reduced 
individually. The reduction included standard bias and flat-field corrections.
The detection and object classification was performed using the software 
package SExtractor (Bertin \& Arnouts 1996).
Instrumental magnitudes were determined with IRAF, using a single aperture, where 
the radius of the aperture is $1.5\times FWHM$. For the deep field, we have stacked
the three exposure after shifting them by $\sim 1-2$ pixels to align them.
Saturation sets in between $V \sim 17$ and 18, depending upon the seeing conditions.

To calibrate our dataset, we chose to use an already standardized dataset of adequate
depth and area coverage. The best match was the data from von Hippel \& Sarajedini
(1998, hereafter vHS98), who have used practically the same $VI$ filters as we have. 
The $I$ filter used by vHS98 is slightly different than the one we have used, since the filter
was replaced in 2001. 
We start the calibration process by using only a constant term in both $V$ and $I$ bands, 
and applying it individually, for each CCD subunit.
We begin with the central-deep field calibrated directly on the vHS98 data. Between 14 and 79
stars are used in each CCD subunit to determine the magnitude offsets, using a $2.5-\sigma$ clipping
of residuals. We obtain a typical scatter of $\sim 0.04$ in the $V$ band and $\sim 0.07$ mags 
in the $I$ band. The larger scatter in the 
$I$ band than in the $V$ band is due to a small color term present in the residuals caused
by the slight difference in the $I$ filter used by the vHS98 observations and ours.
We therefore redetermine the calibration including a linear color term for the $I$ band. 
This is applied for the entire 4 subunit frame using $170$ standards.
The newly obtained colors are checked with the standard values from vHS98 
and no further trends are detected. The linear color term is $0.055\pm0.005$; uncertainties 
in the constant term are of the order of 0.01 mags in both $V$ and $I$ bands.
Once the central-deep field is calibrated, we use it as a standard field for frames that
overlap with it. This calibration process is propagated outward to cover the entire 
area shown in Fig.1. Therefore, our data is on the vHS98 photometric system.
Final magnitudes per object are determined by averaging multiple measurements when available.
The errors are then determined from the scatter of repeated measurements (with $2.5-\sigma$ rejection).
For single measurements, the errors are those calculated by the IRAF routines. Objects with formal IRAF
errors larger than 0.4 magnitudes are not included in the final dataset. 
The positions are determined separately for each CCD 
subunit by tying it into the positions from the P03 catalog. 
The objects are considered matched if they are 
within $0.5\arcsec$ of each other. 

SExtractor's neural network classifier provides a stellarity index value which
can be interpreted as the probability of an object being a point source. Clearly 
the classification is signal-to-noise dependent, and therefore non-uniform across 
our area. We chose to use the stellarity index as given by the $V$ frames,
which have twice the exposure time of the $I$ frames. For multiple measurements, we have also averaged
the indices. In what follows, we will proceed carefully with the interpretation of this
index and eventually use a statistical approach to remove the contribution of galaxies 
to the CMD.
In Figure 2 we show the magnitude errors in each passband as a function of magnitude for the
entire sample (in black), and for the central-deep field (in red). The gray line represents 
a moving median for the entire sample.
The bottom plot shows the stellarity index as a function of magnitude. For the central-deep
field, a reliable classification is obtained down to $V = 22$, where the distribution is
bimodal with peaks near 1 for stars, and near 0 for galaxies. For the entire set,
this limit is more difficult to pinpoint due to the inhomogeneous data;
however at magnitudes fainter than $V =21.5$ the distribution is single peaked.

\section{Color-Magnitude Diagrams: The Overdensity's Distance, 
Metallicity and Age}

We show in Figure 3 the $BV$ CMD from P03 (which covers 0.75 square degrees) and the $VI$ CMD from our 
data (which covers 0.37 square degrees).
We also present the CMDs generated by the Besancon galactic starcount and kinematic model (Robin et al. 2003),
for an area equal to ours, and centered at the location of NGC 188. The top three panels show the 
$BV$ CMDs, while the rest show the $VI$ CMDs for various samples.
The top left panel shows the entire P03 stellar sample as classified in that work. The top middle panel shows
the P03 sample matched with our data set, i.e. restricted to an area equal to that in our study,
and a bright magnitude cut from the OPTIC sample. 
In this panel we show stars as classified in the current work
that achieved a better classification at deeper magnitudes than P03. The top right panel shows 
the Besancon data. A continuous line is drawn to represent the limiting magnitude of the P03 data.
The plots of observed data clearly show an excess of blue stars between $V=19$ and the plate limit, 
that is not seen in the simulated data. Likewise, NGC 188 is not seen in the simulations
and it provides an excellent benchmark to study the newly found overdensity. In the middle row, we show
the $VI$ data for objects with stellar index larger than 0.2 (left), larger than 0.5 (middle) and for the
Besancon data (right). An overdensity is apparent roughly between $V \sim 20$ and 22, and $(V-I) = 0.5$ to 
1.1 when compared to the simulated data. If we apply a more restrictive criterion, i.e., a stellar index
larger than 0.8, we practically eliminate the faint part of the data: essentially NGC 188's main sequence 
ends at $V\sim 21.2$, which is unrealistic. That such a cut causes severe incompleteness of the data, can also 
be seen from Fig. 2, bottom panel.  

The bottom row of the panels shows our data for the central-deep field for the two samples of stellar objects as defined above. 
This field covers an area of only 0.025 square degrees; however a stellar overdensity
appears between $V=21$ and 22 at $(V-I) \sim 1.0$. The scatter in color at these magnitudes due to measurement errors
can be estimated from the scatter in NGC 188's main sequence for single stars at the same magnitude. 
It can be seen that the overdensity's scatter  is comparable to that of NGC 188, indicating that we see a
main-sequence like feature in this central-deep field. 
The accuracy of proper motions from P03 for the few objects in this supposed main sequence feature for 
the central-deep field, is 
unfortunately inadequate for assigning individual membership probabilities at such faint magnitudes.

In Figure 4 we show the objects counts as a function of $V$ magnitude for a sample restricted in color to represent
the region of the overdensity:
$0.5 \le (V-I) \le 1.1$. We show the counts in our $VI$ data for the entire sample (gray), for 
objects with stellar index larger than 0.2 (black), and larger than 0.5 (red). The Besancon star counts are shown with 
a dotted line, and galaxy counts from the Chandra Deep Field South (Gawiser et al. (2006), and see description below)
are shown with a green line. Our data become incomplete at $V\sim 21.2$. However, a clear stellar excess is apparent
between $V = 19$ and 21.6 when compared to the Besancon star counts, assuming the conservative stellar index of 0.5.
Our newly obtained $VI$ data thus confirm the existence of a stellar overdensity
in the background of NGC 188 that was first seen in the P03 data. While the central-deep field hints to a main-sequence
like shape of the overdensity, we further explore this issue with the entire data 
by doing a statistical subtraction of the galaxies and of the Besancon model 
starcounts from our $VI$ CMD. 

Before this however, we 
estimate the overdensity's distance, metallicity and age.
To do so, we use 
the Dartmouth Stellar Evolution Database (Dotter et al. 2008).
First, we fit isochrones by eye in both $BV$ and $VI$ CMDs for NGC 188, as a check and reference.
From our preliminary trials we found that the Dartmouth models that used the 
semi empirical color transformations
of Vandenberg \& Clem (2003) were a better fit than those that used synthetic color transformations.
We have adopted the following well-estimated reddening-related quantities:
$E_{B-V} = 0.09$, $A_{V} = 0.29$, $E_{V-I} = 0.12$ (vHS98, Sarajedini et al. 1999).
For NGC 188, Meibom et al. (2009) have determined a true distance modulus of $11.24\pm0.09$ from an eclipsing binary star.
We adopt this distance modulus and a solar metallicity.
The best-fit isochrone is a 6.5 Gyr model. We note that we obtained poorer fits in the $VI$ CMD when
we used a distance modulus of 11.4 as derived by vHS98, and Sarajedini et al. (1999). 

For the overdensity, we fit isochrones only in the $BV$ data. 
We adopt the reddening determined for NGC 188 (see above), thus assuming that 
there is no reddening between NGC 188 and the overdensity. Since NGC 188 is located 700 pc above the Galactic plane,
this is a realistic assumption.
We explore four metallicity values: [Fe/H] = 0.0, -0.5, -1.0 and -1.5. 
For [Fe/H] = -1.0 and -1.5, the isochrones are moderately $\alpha$-enhanced ([$\alpha$/Fe] = 0.2), to represent a typical 
halo population. We chose the overdensity's supposed main sequence turnoff area at $V \sim 19$ and $(B-V) \sim 0.5$ 
as a reference point through which the theoretical isochrones should pass.
In Figure 5, we show the isochrones overlaid on the $BV$ data, with each panel corresponding
to one metallicity value. The corresponding age and distance modulus range from 2.5 to 12 Gyr, and from 16.0 to 14.8 mag, and
these numbers are specified in each panel. 
From the four cases, the ones that appear to best fit the data are the two intermediate metallicity ones.
We therefore characterize the overdensity to be an intermediate to old population (6 to 10 Gyr) with 
a metallicity between -0.5 and -1.0. The estimated distance is thus between 10.0 and 12.6 kpc. 
We note that this estimation assumes that the overdensity has no depth along the line of sight, which is likely
unrealistic. At a fixed metallicity and age, a given depth of the overdensity will produce a composite main sequence with
a poorly defined turnoff and insufficient curvature to match the main sequence of a single isochrone.
While this may be the case for the $BV$ data, we can not quantify it properly with the data available so far.
On the other hand, we can safely place a lower limit on the distance ($\sim 10$ kpc) which is most important in the
discussion of the plausible orbit of this structure (see Section 5).

In Figure 6 we show the two selected isochrones (in red for [Fe/H] = -0.5, and blue for -1.0) 
overlaid on the $BV$ data (top), and
on the $VI$ data with stellar index $> 0.5$: all area (middle), and the central-deep field (bottom). 
NGC 188's isochrone is also shown (green).
We can see that the isochrones determined from the $BV$ data fit well the $VI$ data.

Given the inability to properly separate stars from galaxies at magnitudes fainter than $V \sim 21$ we aim to 
apply a statistical correction to our $VI$ CMD. For this purpose, we require 
a very deep dataset with well-calibrated photometry and good star/galaxy separation
down to $V \sim 23$. We can then use the CMD of the galaxies in such a dataset to statistically subtract them from
our CMD, assuming galaxies occupy the same region in the CMD. There are numerous multi-band, deep field surveys 
aimed to study field galaxy population and evolution. One such survey is 
the Multiwavelength Survey by Yale-Chile (MUSYC)\footnote{www.astro.yale.edu/MUSYC/} that
is focused on four fields. Of these four, we select two well-known fields: the 
Chandra deep field south (CDFS) at (RA, Dec) = $(53.121\arcdeg, -27.813\arcdeg)$ and (l, b) = $(224\arcdeg, -54\arcdeg)$, and 
the Hubble deep field south (HDFS) at (RA, Dec) = $(338.148\arcdeg, -60.787\arcdeg)$ and 
(l, b) = $(328\arcdeg, -49\arcdeg)$.
The construction of the various catalogs
are described in Gawiser et al. (2006) and Taylor et al. (2009). We work with the catalogs based
on the $BVR$-detection image that lists fluxes in various passbands
including the $VI$ bands which are of interest here, and a stellarity index determined with the same
software package as in our work. We use the fluxes determined from fixed aperture photometry (with an aperture 
correction, see Gawiser et al. 2006) rather than 
other flux estimates more appropriate for galaxies, as we wish to have an observing set resembling
ours as much as possible.  Fluxes are transformed onto the Vega magnitude system according to the transformations
provided in the catalog documentation. These catalogs are complete to $R \sim 25$, and cover an area of
$0.5\arcdeg\times0.5\arcdeg$, or 0.3 square degrees. The star/galaxy separation appears robust to
$V \sim 23$, where the stellarity index has a well-defined bimodal distribution (Gawiser et al. 2006).
We have plotted $VI$ CMDs for stars in both fields and compared them with Besancon Galactic field simulations.
The range in color and the color location of the blue edge and of the red peak representing disk dwarf stars compare
well between the observed and simulated data, assuring us that the photometric calibration is reliable.
We then select galaxies (stellarity index $\le 0.2$) within $V \le 23.0$. We adjust the
$V$ and $(V-I)$ such that both fields have the reddening of the field of NGC 188. Next, we construct Hess diagrams
for the galaxy samples in CDFS and HDFS, and for our entire sample. The galaxy Hess diagrams
are scaled {\it only} by a constant value to adjust for the different spatial area between our data (0.37 sq. deg.) and the CDFS/HDFS 
area (0.3 sq. deg.). Then we subtract each 
galaxy Hess diagram from our data. Here we show only the results from the CDFS, as the other field gives similar 
results. 

The results are shown in Figure 7 for the CDFS. Besides the galaxy-subtracted diagram, we also show  
the subtraction of the Besancon model starcounts from the galaxy-subtracted observed
Hess diagram to highlight the stellar overdensities in this field.
Thus, the top right panel of Figure 7 shows the Hess diagram of our field
with all detected objects included. The top right panel shows the galaxy-subtracted diagram.
The bottom left panel shows the Hess diagram of the galaxy-subtracted and
Besancon-model subtracted data. NGC 188's sequence as well as that of the newly found
overdensity are clearly visible. The bottom right panel shows the galaxy-subtracted Hess diagram with
with the isochrones from Fig. 6 overlaid. 
The faint part of the overdensity's 
main sequence ($V \sim 21.0$ to 22.3) is now better defined. Similarly, 
NGC 188's faint part of the main sequence is more apparent after the galaxy subtraction.
We note that the subtraction over corrects toward the faint end since our data become incomplete
at $V \sim 21.2$, thus reinforcing the reality of the overdensity.
The two isochrones that best characterize this overdensity according to the
$BV$ data are in reasonable agreement with the lower part of the main sequence as seen in the $VI$ data. 
While the region of the turnoff is not very well
defined in the $VI$ data, the blue edge of the $VI$ data is bluer
by $\sim 0.15$ mags
in the magnitude range $V \sim 19.0$ to 20.0, than at brighter magnitudes,
hinting to the presence of a turnoff.

The metallicity and age of the stellar population
of the overdensity determined here agree reasonably well with other determinations for the stellar populations in the Monoceros 
stream such as those of Newberg et al. (2002), or more recently by Ivezi\'{c} et al. (2008). For example, Ivezi\'{c} et al. (2008) estimate a metallicity of -0.95 with a scatter of 0.15 dex from DR6 of the Sloan Digital Sky Survey.


\section{Proper-Motions: Kinematics of the Overdensity}

\subsection{Identifying the Overdensity in Proper-motion Space}

We search the P03 proper-motion data for a kinematically cold, cluster-like structure, other than NGC 188 
to confirm that the main-sequence feature in the CMD is indeed a
stellar stream. Assuming a velocity dispersion of $\sim 10$ km~s$^{-1}$ for stars in a stellar stream 
generated from a massive satellite such as Sagittarius dwarf galaxy (Sgr) for instance (Majewski et al. 2004),
at a distance of 10 kpc, the expected proper-motion dispersion is 0.2 mas~yr$^{-1}$. The proper-motion errors in P03 data
do not reach these values at the faint end of the survey where the overdensity is present. Therefore the proper-motion measuring
uncertainty will dominate the overall proper-motion dispersion. 
For the color range of the overdensity, the proper-motion errors in P03 are:
$ 0.7$ mas~yr$^{-1}$ at $V = 19$, $1.2$ mas~yr$^{-1}$ at $V = 20$, $ 1.7$ mas~yr$^{-1}$ at
$V = 21$, and $ 2.4$ mas~yr$^{-1}$ at 
$V = 21.4$. These are mean values of the formal proper-motion errors from P03.
We thus search for a kinematically cold structure with a dispersion comparable to the size of the proper-motion errors,
which vary substantially at the faint end of the survey.

In Figure 8 we show the relative proper motions from P03.
The top left panel shows the proper-motion distribution of the entire P03 stellar sample. 
The cluster NGC 188 is clearly visible at $\mu_{\alpha}cos\delta = -5.3$ and
$\mu_{\delta} = -0.4 $ mas~yr$^{-1}$ as determined by P03, with a proper-motion scatter $\sim 0.5$ mas~yr$^{-1}$ 
in agreement with formal proper-motion errors. Two proper-motion samples are selected within
the circles drawn. One is located at the approximate center of the overdensity ($(\mu_{\alpha}cos\delta, \mu_{\delta}) = (-3.9, 2.9)$
mas~yr$^{-1}$, on the relative proper-motion system of P03) with a radius of
$1 $ mas~yr$^{-1}$. The second is placed at some arbitrary location to
represent a control Galactic field, and its radius is chosen such that both samples have the same number of objects.
The middle and bottom left panels show the CMD of these two proper-motion selected samples.
The one containing the overdensity shows a well defined main sequence and possibly a subgiant branch, while the control
field does not.
In the top right panel we show our CMD-selection of stars in the overdensity, and another control sample away from the overdensity,
but in the same magnitude range: $V \ge 19$. These two samples have similar numbers of objects. 
The middle and bottom right panels show the proper-motion distributions of the two CMD-selected samples: 
that including the overdensity,
and that excluding it respectively. The middle panel clearly shows a proper-motion clump, while the
bottom panel does not. Therefore, we conclude that the stellar overdensity, not only 
follows a main-sequence like feature in both $BV$ and $VI$ CMDs, but also features a kinematically cold 
population characteristic of tidal stellar streams. 

\subsection{Absolute Proper Motion of the Overdensity}

Next, we want to determine the mean absolute proper motion of the overdensity.
To do so, we work with the $BV$ CMD-selected sample from Fig. 8. 
First, we re-determine the correction to absolute proper motion in this field
using in addition to the 100 galaxies listed by P03, the newly identified ones from our study with stellarity index 
$\le 0.2$. We limit the galaxy
sample to objects brighter than $V = 21.0$ and with proper-motion errors in both coordinates less 
than 2 mas~yr$^{-1}$. We also eliminate
proper-motion outliers, and finally obtain a sample of 143 galaxies. These galaxies average relative 
proper motion, also known as the zero-point correction is 
$\mu_{\alpha}cos\delta = -2.83 \pm 0.18$ mas~yr$^{-1}$ and $\mu_{\delta}
= -1.05 \pm 0.17$ mas~yr$^{-1}$. 
In the P03 study, the stellar proper motions were corrected for magnitude equation relative to cluster members
with $V = 16.0$ (their Fig. 2). 
However P03 chose not to apply this rather strong magnitude equation to the galaxies, 
as they showed no detectable trend in their proper motions with magnitude. 
For the purpose of determining the absolute proper motion of the overdensity, we have chosen
 to work with the data from P03 but {\it uncorrected} for magnitude equation. Our choice
is justified by the fact that in P03 the magnitude equation in proper motions is poorly
constrained at $V > 19$. At these magnitudes it was difficult to select reliable cluster members
from the $BV$ CMD which were used to derive the magnitude equation.
We inspect the proper motions of galaxies (including the fainter ones identified in this work) as a function of
magnitude, as well as the proper motions of the CMD-selected overdensity stars as a function of magnitude.
We detect a very small trend with magnitude in $\mu_{\delta}$.
This trend is about 0.2 mas~yr$^{-1}$mag$^{-1}$ and is similar for galaxies and for stars in the overdensity.
The average magnitude for galaxies is $V = 19.9$ and that of stars in the overdensity is $V = 20.4$.
The combination of the small difference between the mean magnitude of the two samples and their similar 
magnitude trend allows us to safely proceed with no magnitude equation correction to either galaxies or overdensity stars.

To determine the mean absolute proper motion of the overdensity,
we apply the well-known techniques developed
for open-cluster studies that separate the cluster population from the Galactic field population.
Here we follow the procedure described in Girard et al. (1989) where both the cluster and the field
have their intrinsic proper-motion distributions described by Gaussians. 
Typically, observed proper-motion distributions are built in each coordinate and fitted with 
a sum of two Gaussians representing the field and the cluster.  
The width of the Gaussian distribution for the cluster reflects
the intrinsic dispersion of the cluster, the measuring uncertainty of the proper motions and a contribution
due to the smoothing process done when constructing the observed distributions (for details see Girard et al. 1989, Dinescu 
et al. 1996). 
The analysis is simple when the proper-motion errors have approximately the same value. However,
in our case the proper-motion errors vary significantly with magnitude. For such cases,
as shown by Girard et al. (1989) it is better to construct  an error
function from the formal proper-motion errors. This error function is convolved with the function describing
the intrinsic proper-motion distribution of the cluster (and field), which in this case is taken to be a Gaussian,
and then fitted to the observed proper-motion distribution. The parameters from the fit  are
the number of cluster stars, the mean proper motion of the cluster and of the field and the intrinsic 
dispersion of the cluster and of the field. We have tried various fits for our sample by experimenting
with the size of the formal proper-motion errors (i.e., increasing them by $25\%$) and with the
number of stars in the overdensity. In all these fits, the mean motion of the overdensity appears
with a robust determination and a variation of $\sim 0.2$ mas~yr$^{-1}$.  The
mean absolute proper motion of the overdensity is 
$\mu_{\alpha}cos\delta = -1.48\pm0.27$ and
$\mu_{\delta} = 1.98\pm0.26$ mas~yr$^{-1}$, where the final error includes the contribution from
the variation in the fit and that from the absolute proper-motion zero point. From the fits, we also
obtain approximately 200 stars as members of the overdensity from a total of 482 stars in the CMD-selected sample.
In Galactic coordinates, the proper motion is $\mu_l~cos~b = -1.40\pm0.27$ and $\mu_b = 2.04\pm0.26$ 
mas~yr$^{-1}$.

\section{Discussion}

\subsection{The Overdensity as Part of the Monoceros Stream}

The overdensity is located at (l,b) = $(122.8\arcdeg, 22.5\arcdeg)$. 
In what follows, we will work with two values for the distance to the overdensity
determined in Section 3: 10.0 and 12.6 kpc. 
For 10 kpc, the overdensity is at $(X,Y,Z) = (13.0,7.8,3.8)$ kpc, 
while for 12.6 kpc it is at $(X,Y,Z) = (14.3,9.8,4.8)$ kpc, 
where the Sun is at  $(X,Y,Z) = (8,0,0)$ kpc.
Besides the properties of the stellar population, another hint that this overdensity 
may belong to the Monoceros stream is its Galactic location.
Of the numerous studies that have sampled this structure, here we mention those that have found
positive detections near our region: Rocha-Pinto et al. (2003) analyze M giants from 2MASS to map out
this structure; indeed they find positive detections at (l,b) $ \sim (130\arcdeg, 26\arcdeg)$ with 
an inferred heliocentric distance between 9 and 13 kpc. Conn et al. (2005) trace the main
sequence of Mon using the Wide Field Camera on the Isaac Newton Telescope. They find a
positive detection at (l,b) = $(118\arcdeg, 16\arcdeg)$ and at a heliocentric distance of $\sim 12$ kpc.
Momany et al. (2006) investigate an alternative explanation of the overdensity at the Galactic location of 
the Conn et al. detection, namely the flared disk. Their model of the warped and flared disk is based on 
2MASS red clump and red giant stars.
For this Galactic location, Momany et al. provide a map of the scale height of the flared disk as a function of distance from the
Galactic plane and from the Galactic center. This map is the 
closest in direction, of those presented, to our NGC 188 field. 
In this map our overdensity, for both distance determinations,
is located beyond the contour corresponding to $3\times$ the scale height of the flared disk. Therefore,
its position alone makes the overdensity unlikely to belong to the flared disk as recently modeled by 
Momany et al. (2006).

In order to provide further evidence for this tentative identification,
we compare our distance estimation and proper motion to the predictions of the
Pe\~{n}arrubia et al. (2005, hereafter P05) model for the Monoceros stream. This model describes
the disruption of a satellite on a prograde, low inclination, low eccentricity orbit and is
constrained by spatial distribution, distance estimates and radial velocities in overdensities
mapped above and below the Galactic plane and between $l = 110\arcdeg - 240\arcdeg$.
In Figure 9, we show the spatial distribution of the debris from the P05 model (top panel). The location
of our field is marked (red symbol), and a sample of model particles selected within $4\arcdeg\times4\arcdeg$ centered
around our field are also shown (black squares). The middle and bottom panels show the
proper motions in Galactic longitude and latitude as a function of distance for the entire 
model (grey), for the particle sample coincident with our field (squares) and for our field (red). The model
indicates that there are two distance groups, one at $\sim 9-12$ kpc, and the other at $17-23$ kpc.
The near group, which corresponds well with our data divides into two groups according to the 
proper motion in latitude: one group moves downward, toward the Galactic plane, the other, away from
the plane. Our data fits very well with the group moving away from the plane. Located at a distance of
3.8/4.8 kpc from the plane, stars in this overdensity will continue to get further away from the plane as they
proceed in a prograde sense. 
We have found no evidence in the proper-motion data of the other group postulated by the model, i.e. the one moving
toward the plane. Likewise, we find no evidence in our data of the more distant group at 20 kpc, which may be 
due to the magnitude limit of our observations.

\subsection{Plausible Orbits}

As we do not have a measurement of the radial velocity (RV) for stars in the overdensity, we explore a range of
plausible values in order to have a better understanding of the likely orbits constrained by 
the absolute proper motion and by the distance estimate.
From the Besancon model, stars in the color ($B-V < 1.0$) and magnitude ($17 < V < 22$) ranges 
occupied by the overdensity have a heliocentric RV distribution centered at $\sim -70$ km~s$^{-1}$, with
a FWHM of 140 km~s$^{-1}$, and a tail toward negative values of $\sim -300$ km~s$^{-1}$. We therefore choose the
following RV values to be explored: 0, -70, -140 and -300 km~s$^{-1}$ and two extreme values of 100 and -400 km~s$^{-1}$.
In Table 1 we list the adopted heliocentric RV, and the resulting velocity components in cylindrical coordinates $(\Pi, \Theta, W)$. 
We also list the maximum distance from the plane and the eccentricity of the respective orbits. The adopted peculiar velocity
of the Sun is from Dehnen \& Binney (1998). The velocity of the local standard of rest is $(\Pi, \Theta, W)$ = (0, 220, 0) km~s$^{-1}$.
We have integrated the orbit in the Johnston et al. (1995) Galactic potential.

\begin{table}[htb]
\caption{Plausible Orbits for the Newly Found Structure}
\begin{tabular}{rrrrrrrrrrrrr}
\tableline
\\
 & \multicolumn{5}{c}{d = 10.0 kpc} & & & \multicolumn{5}{c}{d = 12.6 kpc} \\
\tableline
\multicolumn{1}{c}{$V_r$} & \multicolumn{1}{c}{$\Pi$} & \multicolumn{1}{c}{$\Theta$} & \multicolumn{1}{c}{W} & \multicolumn{1}{c}{$z_{max}$} & \multicolumn{1}{c}{e} &  & & \multicolumn{1}{c}{$\Pi$} & \multicolumn{1}{c}{$\Theta$} & \multicolumn{1}{c}{W} & \multicolumn{1}{c}{$z_{max}$} & \multicolumn{1}{c}{e} \\
\multicolumn{1}{c}{(km~s$^{-1}$)} & \multicolumn{3}{c}{(km~s$^{-1}$)}  & \multicolumn{1}{c}{(kpc)} & & && \multicolumn{3}{c}{(km~s$^{-1}$)}
& \multicolumn{1}{c}{(kpc)} & \\
\tableline
\\
0 & 44 & 241 & 99 & 7.3 & 0.27  &&& 43 & 250 & 123  & 11.0 & 0.34 \\
-70 & -14 & 212 & 72 & 5.8 & 0.03 &&& -17 & 225 & 96 & 9.0 & 0.13 \\
-140 & -72 & 184 & 46 & 5.0 & 0.23 &&& -77  & 200 & 69  & 8.0  & 0.19 \\
-300 &  -205 & 118  & -16 & 4.2 & 0.70 &&& -213 & 142 & 8 & 7.6  & 0.66 \\ \\
 100 & 126 & 282 & 137 & 11.7 & 0.59 &&& 128 & 286 & 161 & 18.2 & 0.64 \\
-400 & -288 & 77 & -54 & 5.2 & 0.89 &&& -298 & 106 & -30 & 9.1 & 0.86 \\
\tableline
\end{tabular}
\end{table}

In all cases explored here that cover the entire plausible RV-range, the orbit is prograde.
Also, the maximum distance above the plane is rather high, thus making the orbits unlikely
to represent material in the warped or flared disk as described by Momany et al. (2006) for instance. 
For the 12.6 kpc distance estimate, the orbits are even more inconsistent with a thick-disk origin of this overdensity than for the 10-kpc distance estimate, as the material travels quite far from the Galactic plane.

The orbit with the lowest height above the plane (i.e. 4.2 kpc, for the 
10.0 kpc distance estimate) has a highly eccentric orbit, again inconsistent
with material belonging to the thick disk. 
To achieve an orbit more confined to the Galactic plane, for instance
within $\pm 3$ kpc from the Galactic plane 
we require a distance to the overdensity of 7 kpc for our measured absolute proper motion.
This implies a brighter turnoff by $\sim 0.8$ mag, which the $BV$ data do not support.

However, some of the orbits explored here (RV = 0 to -140 km~s$^{-1}$) 
are quite compatible with the one proposed by P05 for the model of the Monoceros stream.
Therefore, the orbits discussed in combination with the properties of the stellar population of the newly
found overdensity argue in favor of its membership to the Monoceros stream.

\section{Summary}

We present evidence for a stellar overdensity in the background of the open cluster NGC 188. 
By fitting the main-sequence of the overdensity with isochrones in the $BV$ 
CMD
we constrain its stellar population to be  
between 6 and 10 Gyr and with a metallicity [Fe/H] = -0.5 to -1.0. The 
corresponding distance from Sun is between 10 and 12.6 kpc.
In proper-motion space, it is shown to be a kinematically cold population.
The absolute proper motion, distance and Galactic location of the
overdensity agree reasonably well with the predictions of the P05 model for the formation of the Monoceros stream.
We also derive plausible orbits by adopting a range of RVs and show that the orbits are incompatible 
with material belonging to the warp or flare of the disk.

\acknowledgments

We thank Jorge Pe\~{n}arrubia for making available to us the simulated data for the Monoceros stream model.
Financial support from National Science Foundation through grant AST 04-06884 for this research is acknowledged. 
I. Platais gratefully acknowledges support from the NSF through grant AST 09-08114 to Johns Hopkins University.

{\it Facilities:} \facility{WIYN 3.5m}.

\clearpage

\begin{figure}
\includegraphics[angle=-90,scale=.70]{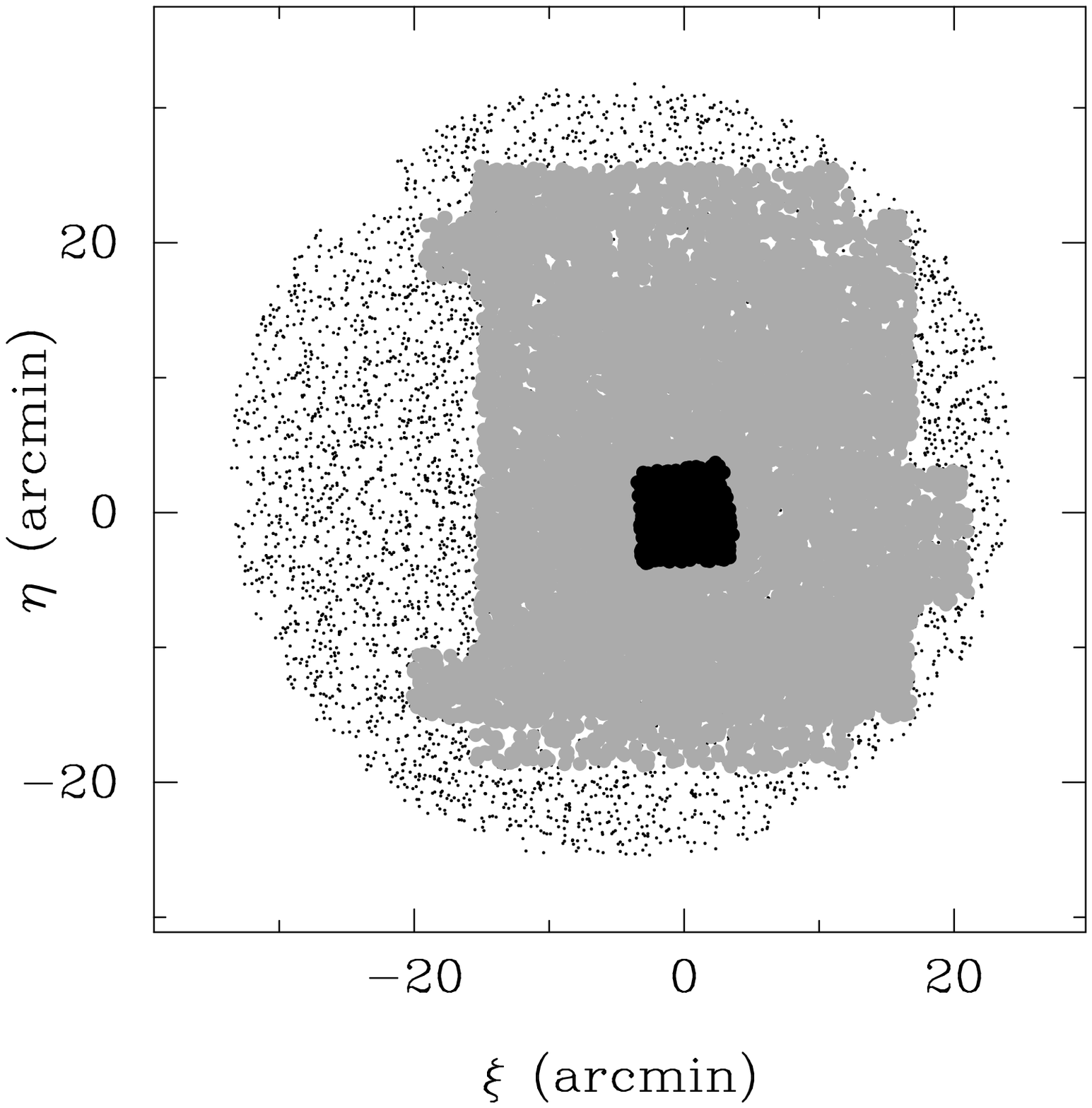}
\caption{Area coverage of various datasets used in this paper: the P03 proper-motion data (black dots),
the OPTIC data (grey), and the Von Hippel \& Sarajedini (1998) calibrating data set (black filled circles).}
\end{figure}

\clearpage

\begin{figure}
\includegraphics[scale=.70]{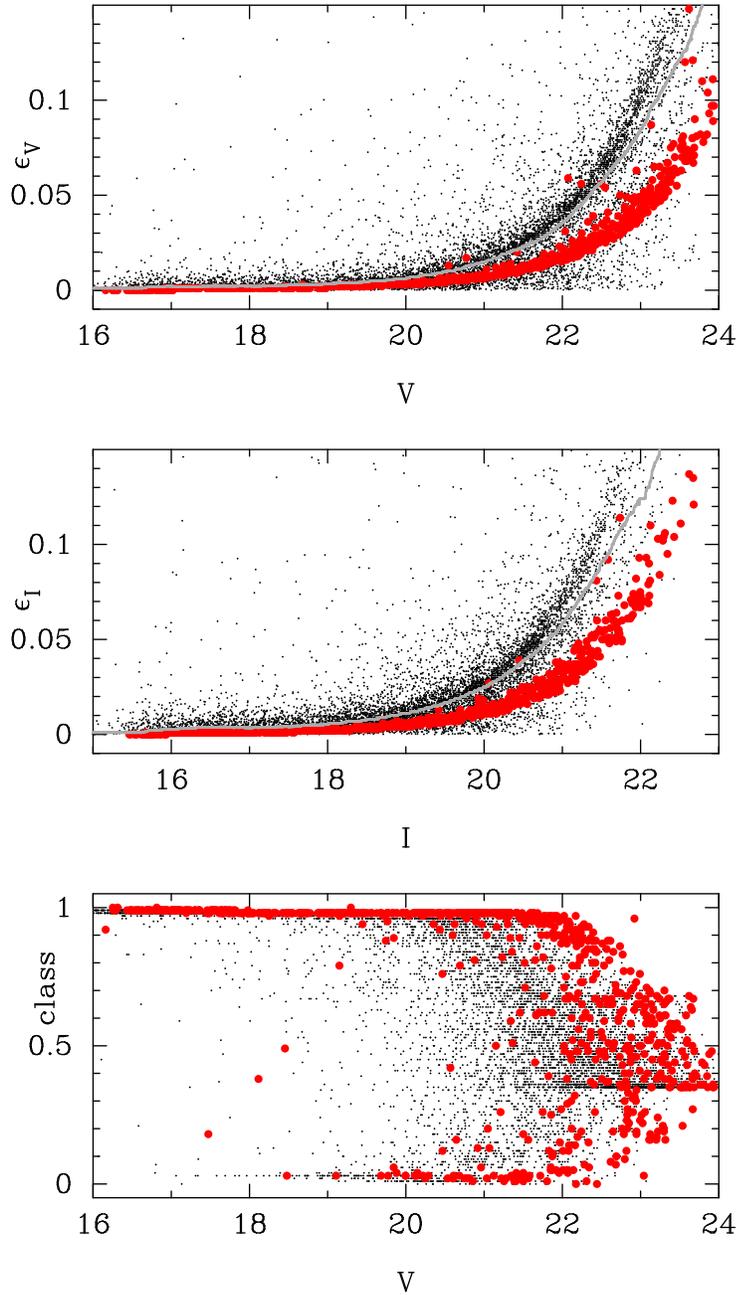}
\caption{Magnitude errors in $V$ (top) and $I$ (middle), and  stellar index
as a function of magnitude (bottom). The gray line shows a moving median for the entire sample. The red
symbols show the data for the central-deep field.}
\end{figure}

\begin{figure}
\includegraphics[scale=.80]{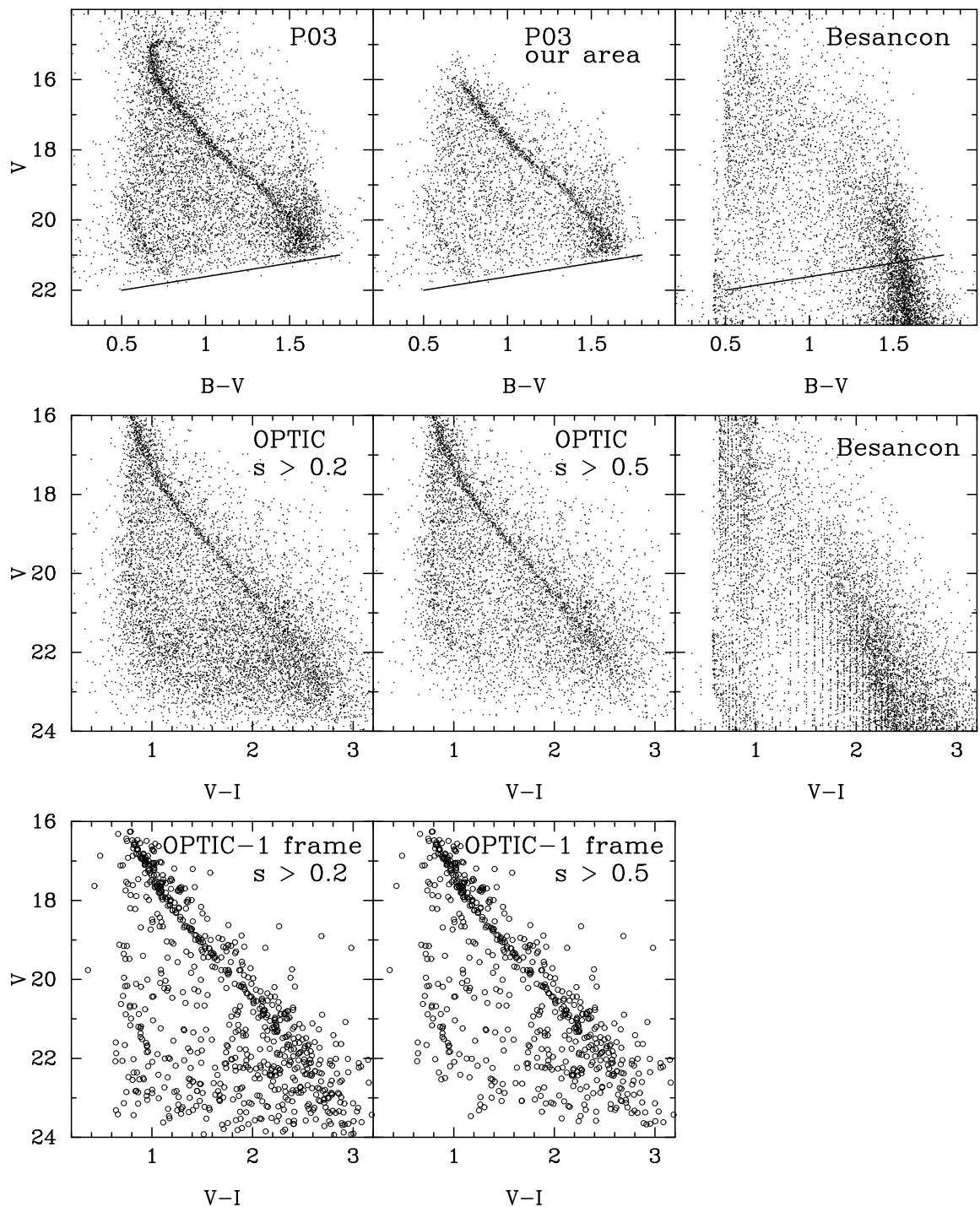}
\caption{Top panels: $BV$ CMDs of stars in P03 (left), in P03 data restricted to our area (middle), and as predicted by the 
Besancon model (right) in the same area as that covered by our study. The middle panels show
the $VI$ CMDs of our stellar sample where the stellar index is $ > 0.2$ (left), $ > 0.5$ (middle), and of
the Besancon model (right). The bottom panels show the CMDs for stars in the central-deep field, where
the stellar index is $ > 0.2$ (left) and $ > 0.5$ (middle).}
\end{figure}

\begin{figure}
\includegraphics[scale=1.0]{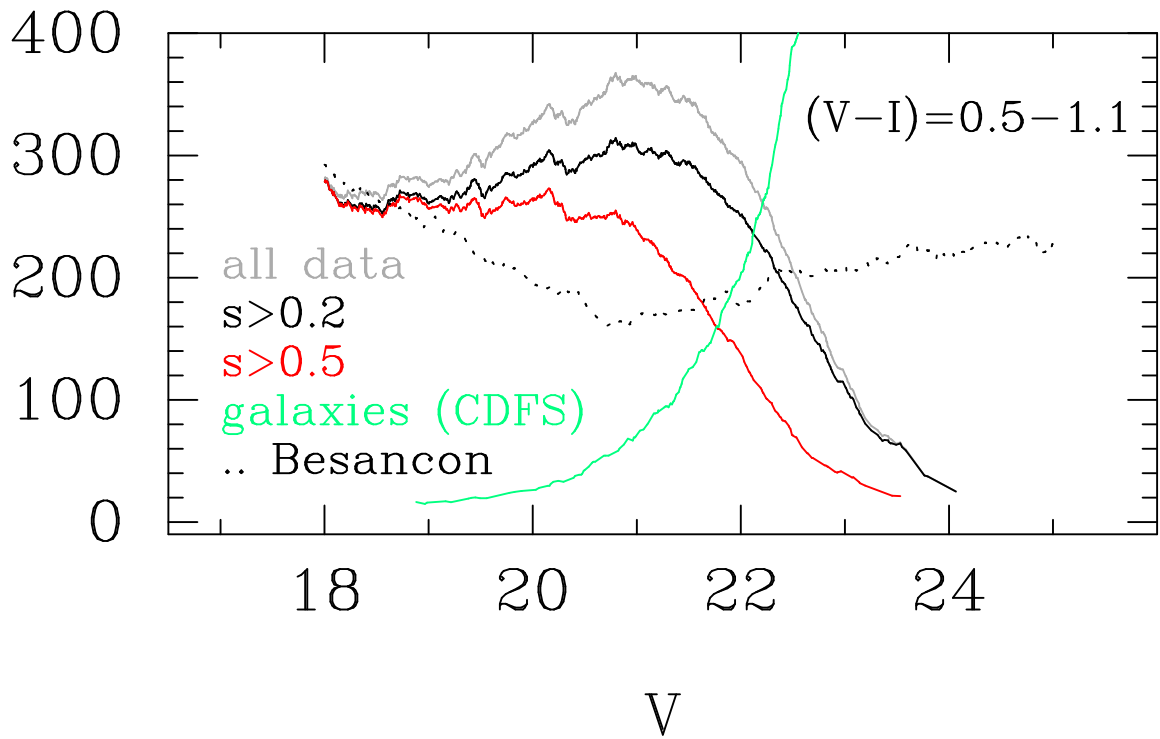}
\caption{Object counts for $0.5 \le (V-I) \le 1.1$ as a function of $V$ magnitude for our data: 
all sample - gray, stellar index $> 0.2$ - black,
stellar index $>0.5$ - red, for the Besancon model - dotted, and for galaxies(CDFS) - green.} 
\end{figure}

\begin{figure}
\includegraphics[angle=-90.0,scale=.7]{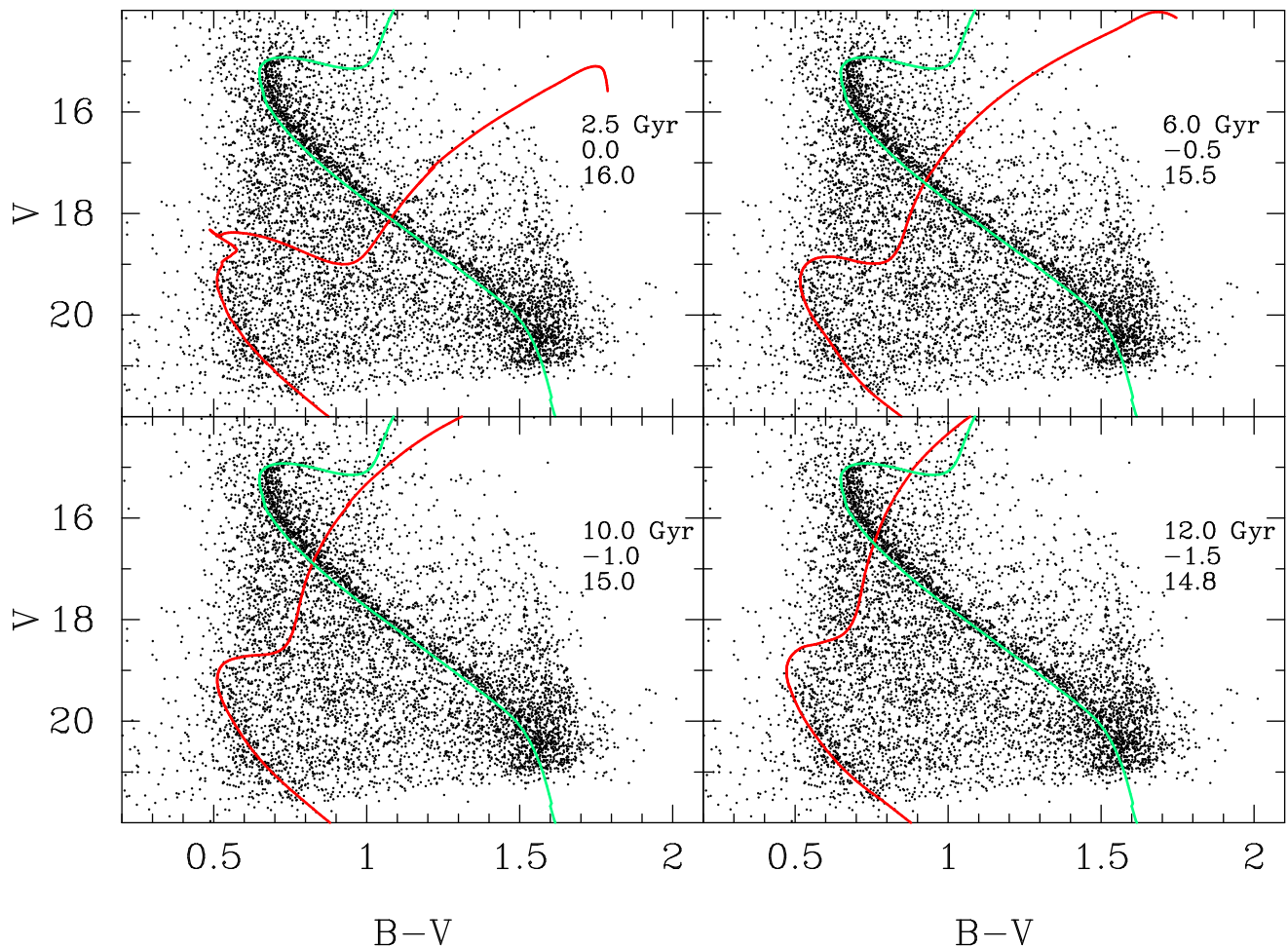}
\caption{Dartmouth isochrones fit to the $BV$ CMD data. For NGC 188 we used a 6.5 Gyr model of solar metallicity (green).
The adopted reddening is $E_{B-V} = 0.09$.
For the overdensity we show isochrones (red) of four metallicity values. These values together with the age and distance modulus
are specified in each panel.}
\end{figure}

\begin{figure}
\includegraphics[scale=.7]{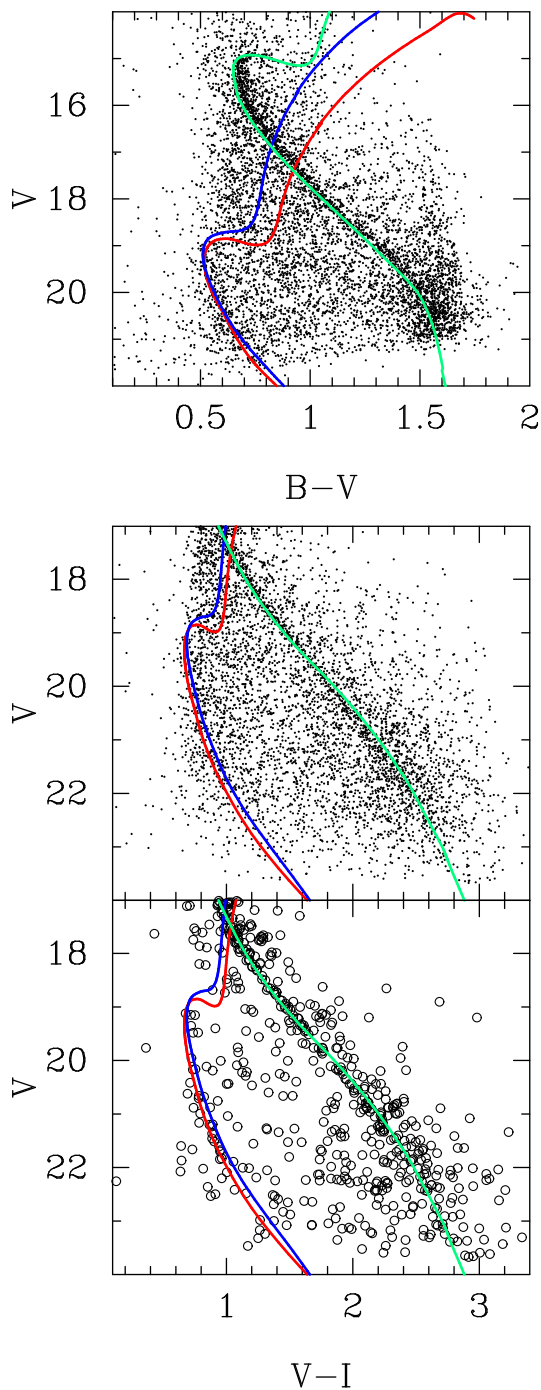}
\caption{Dartmouth isochrones overlaid on the $BV$ CMD data (top) and on the $VI$ 
data (middle) and central-deep field (bottom). In the $VI$ data objects with stellar index $>0.5$ are shown.
For the overdensity, we use 6 Gyr models with [Fe/H] = -0.5 (red), and 10 Gyr with [Fe/H] = -1.0 (blue),
with distance moduli of 15.5 and 15.0 respectively.}
\end{figure}

\begin{figure}
\includegraphics[scale=.8]{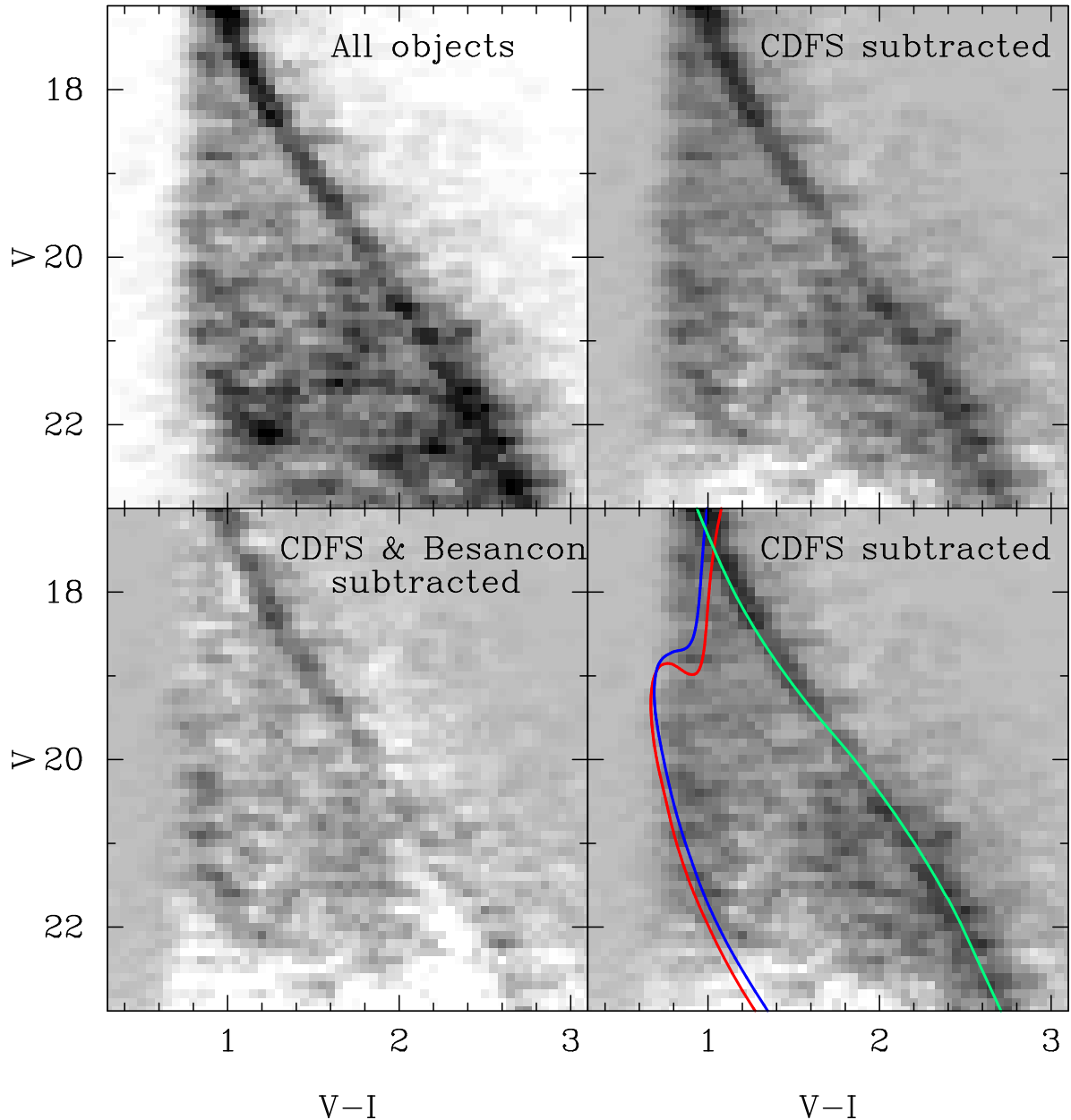}
\caption{Statistical subtraction of galaxies and of the Besancon model from the observed CMD.
The galaxies are from the MUSYC survey for the CDFS area. Our entire sample is shown in the top-left panel,
the galaxy-subtracted one is in the top-right panel. The bottom-left panel shows the galaxies and 
Besancon starcount data subtracted from the observed CMD. In the bottom-right panel we show
 the galaxy-subtracted sample with the isochrones from Fig. 6 overlaid.}
\end{figure}

\begin{figure}
\includegraphics[scale=.75]{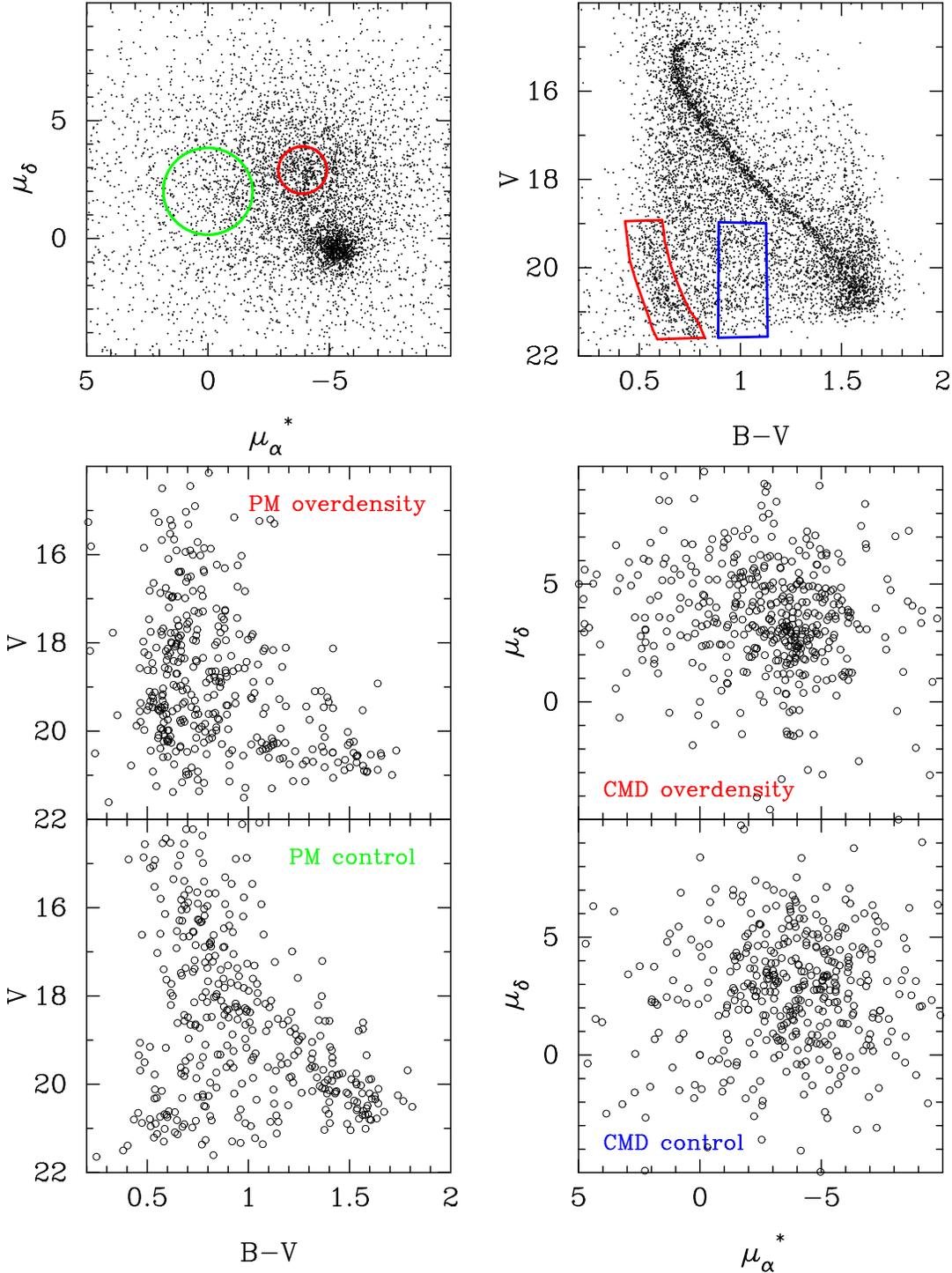}
\caption{Relative proper-motion diagram showing all stars in the P03 data (top-left). The obvious concentration of stars represents NGC 188. 
Two proper-motion samples are selected from this diagram within the circles shown. The circles include samples of same number of objects.
One is centered on the overdensity region, the other on a control region.  The middle and bottom left panels show the CMD for the proper-motion selected samples. 
The top-right panel shows the CMD-selection of
two samples: one representing the overdensity, and the other
a control Galactic field.
The middle and bottom right panels show the proper-motion diagrams of these two samples. 
The overdensity sample shows a clear proper-motion clump, while the other sample does not.}
\end{figure}

\begin{figure}
\includegraphics[scale=.8]{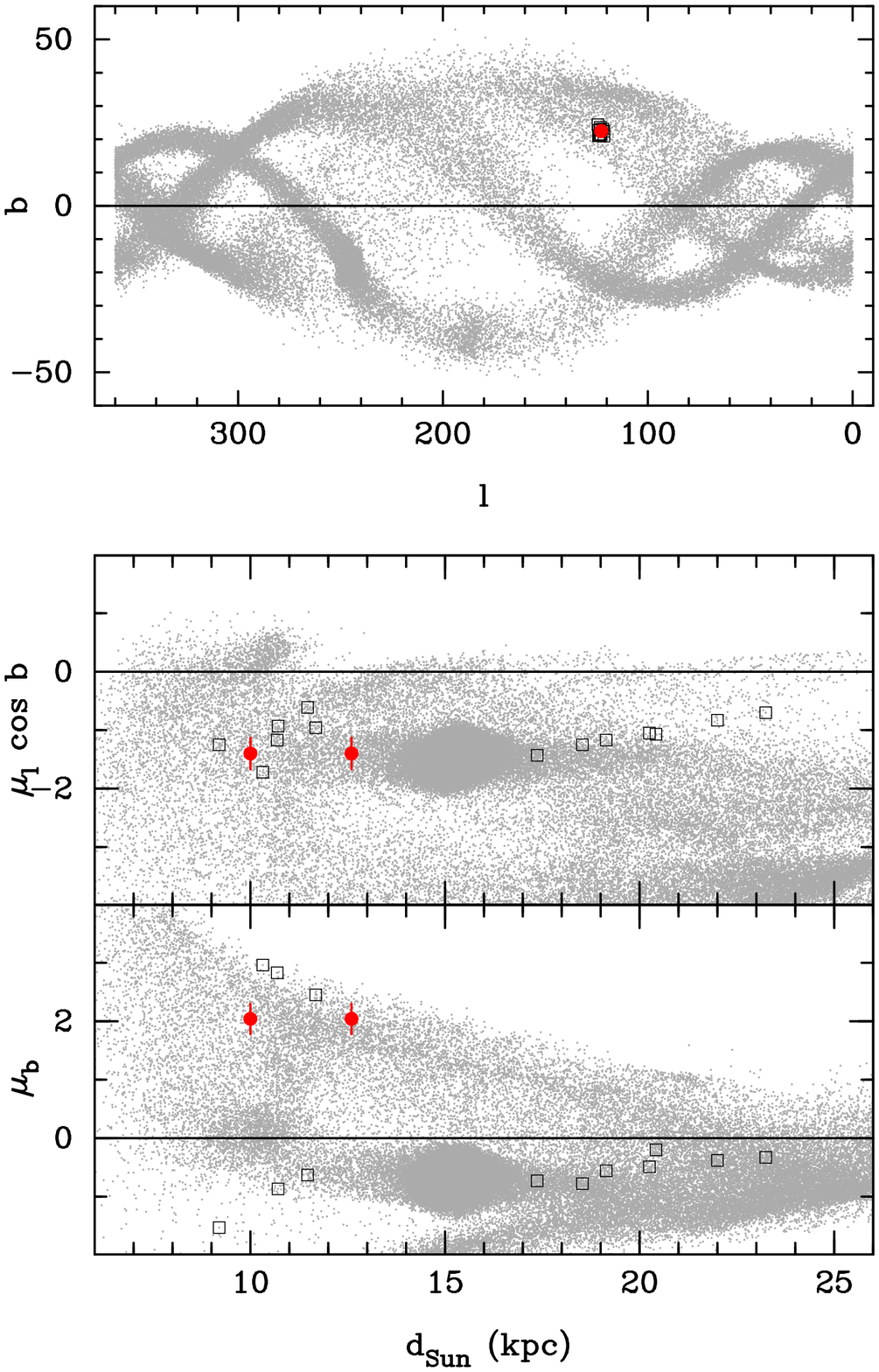}
\caption{Comparison of the Pe\~{n}arrubia et al. (2005) model for the disruption of the progenitor satellite 
of the Mon ring with the proper-motion data. The top panel shows the distribution in Galactic
coordinates of the model (gray), our field for two distance estimates (red), and a sample from the model representing our field (black
open squares). The middle and bottom panels show the proper motions in Galactic coordinates
as a function of distance from the Sun. Again, the black open squares show the
predicted kinematics at a pointing corresponding to our field. 
The mean proper motion of the overdensity and its distance agree
with predictions from the model for the more nearby part of the stream that moves away 
from the Galactic plane.}
\end{figure}


\begin{references}
\reference{} Bertin, E. \& Arnouts, S. 1996, \aaps, 117, 393
\reference{} Casetti-Dinescu, D. I., Majewski, S. R., Girard, T. M., Carlin, J. L., van Altena, W. F., Patterson, R. J., \& Law, D. R. 2006, \aj, 132, 2082
\reference{} Casetti-Dinescu, D. I., Carlin, J. L., Girard, T. M., Majewski, S. R., Pe\~{n}arrubia, J. \& Patterson, R. J. 2008, \aj, 135, 2013 
\reference{} Casetti-Dinescu, D. I., Girard, T. M., Majewski, S. R., Vivas, A. K., Wilhelm, R., Carlin, J. L., Beers, T. C., 
\& van Altena, W. F. 2009, in press
\reference{} Conn, B. C, Lewis, G. F., Irwin, M. J., Ibata, R. A., Ferguson, A. M. N., Tanvir, N., \& Irwin, J. M. 2005,
\mnras, 362, 475
\reference{} Conn, B. C., et al. 2007, \mnras, 376, 939
\reference{} Conn, B. C., Lane, R. R., Geraint, L. F., Irwin, M. J., Ibata, R. A., Martin, N. F., 
Bellazzini, M., \& Tuntsov, A. V.  2008, \mnras, 390, 1388
\reference{} Dehnen, W \& Binney, J. 1998, \mnras, 294, 429
\reference{} Dinescu, D. I., Girard, T. M., van Altena, W. F., Yang, T-G., Lee, Y-W. 1996, \aj, 111, 1205
\reference{} Dinescu, D. I., Girard, T. M., \& van Altena, W. F. 1999, \aj, 117, 1792
\reference{} Dinescu, D. I., Majewski, S. R., Girard, T. M., \& Cudworth, K. M. 2000, \aj, 120, 1892
\reference{} Dinescu, D. I., Girard, T. M., van Altena, W. F., \& L\'{o}pez, C. E. 2005, \apj, 618, L25 
\reference{} Dotter, A., Chaboyer, B., Jevremovi\'{c}, D., Kostov, V., Baron, E. \& Ferguson, J. W. 2008, \apjs, 178, 89
\reference{} Gawiser, E. et al. 2006, \apjs, 162, 1
\reference{} Girard, T. M., Grundy, W., L\'{o}pez, C. E. \& van Altena, W. F. 1989, \aj, 98, 227
\reference{} Ibata, R., Gilmore, G., Irwin, M. J. 1994, Nature, 370, 194
\reference{} Ibata, R., Lewis, G. F., Irwin, M. J., Totten, E., Quinn, T. 2001, \apj, 551, 294
\reference{} Ivezi\'{c} Z. et al. 2008, \apj, 684, 287
\reference{} Jacoby, G., Tonry, J. L., Burke, B. E., Claver, C. F., Starr, B., Saha, A., 
Luppino, G. A., \& Harmer, C. 2002, SPIE, 4836, 217
\reference{} Johnston, K. V., Spergel, D. N., \& Hernquist, L. 1995, \apj, 451, 598
\reference{} Juri\'{c}, M. et al. 2008, \apj, 673, 864 
\reference{} Majewski, S. R., Skrutskie, M. F., Weinberg, M. D., \& Ostheimer, J. C., 2003, \apj, 599, 1115
\reference{} Majewski, S. R., Kunkel, W. E., Law, D. R., Patterson, R. J., Polak, A., Rocha-Pinto, H. J., Crane. J. D.,
Frinchaboy, P., Hummels, C. B., Johnston, K. V., Rhee, J., Skrutskie, M. F., \& Weinberg, M. 2004, \aj, 128, 245
\reference{} Meibom, S., Grundhal, F., Clausen, J. V., Mathieu, R. D., Frandsen, S., Pigulski, A., Narwid, A., Steslicki, M., \& Lefever, K. 2009, \aj, 137, 5086
\reference{} Momany, Y., Zaggia, S., Gilmore, G., Piotto, G., Carraro, G., Bedin, L. R., \& De Angeli, F. 2006, \aap, 451, 515
\reference{} Newberg, H. et al. 2002, \apj, 569, 245
\reference{} Pe\~{n}arrubia, J., Mart\'{i}nez-Delgado, D., Rix, H. W., G\'{o}mez-Flechoso, M. A. (P05)
\reference{} Platais, I., Kozhurina-Platais, V., Mathieu, R. D., Girard, T. M., \& van Altena, W. F. 2003, \aj, 126, 2922 (P03)
\reference{} Robin, A. C., Reyl\'{e}, C., Derri\`{e}re, S., \& Picaud, S. 2003, \aap, 409, 523
\reference{} Rocha-Pinto, H. J., Majewski, S. R., Skrutskie, S. R., \& Crane, J. D. 2003, \apj, 594, 115
\reference{} Taylor, E. N., et al. 2009, \apjs, 183, 295
\reference{} Tonry, J. L., Burke, B. E., \& Schechter, P. L. 1997, \pasp, 109, 1154
\reference{} VandenBerg, D. A., \& Clem, J. L. 2003, \aj, 126, 778
\reference{} Vivas, A. K., \& Zinn, R. 2003, Mem. Soc. Astron. Italiana, 74, 928
\reference{} von Hippel, T. \& Sarajedini, A. 1998, \aj, 116, 1789
\reference{} Yanny, B. et al. 2003, \apj, 588, 824
\end{references}
\end{document}